# Detection and Analysis of Emotion From Speech Signals

Assel Davletcharova[a], Sherin Sugathan[b], Bibia Abraham[c], Alex Pappachen James[a,b]

[a]Department of Electrical & Electronic Engineering, Nazarbayev University, Kazakhstan.
[b]Enview Research & Development Labs, Trivandrum, India.
[c]Medical College Hospital, Kannur, India.

## Abstract

Recognizing emotion from speech has become one the active research themes in speech processing and in applications based on human-computer interaction. This paper conducts an experimental study on recognizing emotions from human speech. The emotions considered for the experiments include neutral, anger, joy and sadness. The distinuishability of emotional features in speech were studied first followed by emotion classification performed on a custom dataset. The classification was performed for different classifiers. One of the main feature attribute considered in the prepared dataset was the peak-to-peak distance obtained from the graphical representation of the speech signals. After performing the classification tests on a dataset formed from 30 different subjects, it was found that for getting better accuracy, one should consider the data collected from one person rather than considering the data from a group of people.

## 1. Introduction

Emotion classification[1] is one of the most challenging tasks in a speech signal processing domain. The problem of speaker or speech recognition becomes relatively an easier one when compared with recognizing emotion from speech. Sound signal is one of the main medium of communication and it can be processed to recognize the speaker, speech or even emotion. The basic principle behind emotion recognition lies with analysing the acoustic difference that occurs when uttering the same thing under different emotional situations. In addition to the features corresponding to the speaker and/or the speech, the sound signals do have some features that represents the emotional state of the speaker. The paper addresses the problem of emotion classification for human speech. The study is aimed at exploring dependencies the nature of utterance have with the human emotional state. Since the emotions have a direct influence on the nervous system, the heart rate also is affected by them. So the heart rate of a person can also be measured to get information about the emotional status of person[2, 3]. It is interesting to note that the speech signals are also a representative of the heart rate of the speaker since the heart rate also affects the speech. The work in [4] says that if there is a negative stimuli that causes negative emotion the heart rate decelerate more actively than when there is positive stimuli [4].

---

✩Corresponding Author: Email: apj@ieee.org

Currently, speech signal processing is widely used in many applications including non-contact medical diagnosis, remote observation of patients, human-computer interaction (HCI) etc. Speech signals can be analysed to detect heart rate, which will naturally enable a remote diagnosis of a patient by only using audio data. Other medical applications include diagnosis of nasopharynx and vocal tract for any abnormalities.

2. Background

In [5], it is argued that heart rate variation (HRV) has some dependency on emotional state of person. The results of the study revealed that the emotional stress can cause death due to heart disorders like acute myocardial infarction. It can also make prediction on risk of developing the hypertension. However, it was proved that positive emotions can be beneficial in treatment of that hypertension by causing alterations in HRV.

The idea proposed in [6] used speech parameters for detection and analysis of the vocal fold pathology. There were developed algorithms for speech signal processing in order to create a model for characterizing healthy and pathology conditions of human vocal folds. The methodology was based on extraction of the separate speech signal components from both healthy and assumed pathology conditions. The iterative maximum likelihood (ML) estimation was applied for solving that problem.

In [7], an algorithm was developed for the detection of hypernasal resonance. This approach is important because it provides noninvasive contactless interaction with patient, hence maximizing the accurate detection of speech due to the naturalness of speaking in comfort conditions without extra devices on the face and body.

The work in [2, 3] studies the dependency of the physiology with human emotions in application to HCI. The approach uses physiological characteristics like temperature and electro dermal activity as an input for emotion recognition. The results indicate the possibility of developing an emotional relationship between humans and computers which enables the development of a human-friendly personal robot [2, 3].

In [4], they proposed a method of measuring the heart rate of a patient sitting on a chair. The method was based on using of electro-mechanical film and traditional ear lobe photo-plethysmo graph (PPG) embedded into the chair. The experiment was conducted on twenty four participants in order to demonstrate whether human emotions can be directed to computer in the same way as to society.

In [8], they conducted an experiment of speech therapy through telemedicine technology with a group of patients. The experiment showed that interaction among patients through speech accelerates their recovery and positively influence on the quality of life [8]. The same could be improved by monitoring heart rate detected from the speech of participants and evaluate the progress of recovering as well as prevent any risks related to the heart failure.

In [9], they introduced a remote detection of the Body Mass Index from the speech signal. The novel approach was designed for remote monitoring the patients' weight in order to control the risks of diseases and death that are resulted from underweighting or in opposite from overweighting. The importance of the telemedicine was also discussed in this study, showing its benefits from different aspects such as comfort, low cost, time efficiency, etc [9].

The work in [10] proposed an innovative approach of using mobile phones for measuring the heart rate continuously. The design consisted of three sub-systems; the first one records the signals and performs an offline analysis of the heart rate; the second system provides support for remote real time monitoring of the detected electrocardiogram (ECG) signal by sending the data to the medical centre or doctor through certain communication media; the third system performs a local real time classification of collected data. The main advantage of this design is the support for mobility of both patient and doctor[10]. This design can be improved by applying speech signal recording for evaluating the heart rate instead of using ECG sensor that sends the signal to the mobile phone via bluetooth connection.

3. Methodology

For studying the basic nature of features in speech under different emotional situations, we used data from three subjects. As part of the data collection, we recorded the voice of three different female subjects. The subjects were asked to express certain emotions when the their speech was recorded. The subjects were Russians and they spoke Russian words under different emotional states. A mobile phone was used to record the speech and was kept at a distance about 15cms away from the mouth. The experiments were conducted in an ordinary bedroom having an area of 25m$^2$. For extracting features from the recorded speech segments, MATLAB functions [11] were used.

The analysis of the recorded speech signals were done in a MATLAB environment which provides several graphical visualizations for analysing a signal. In Figure.1(a), a power spectrum of the speech signal is shown which indicates the peak value using a dark red color. In Figure.1(b), a linear graphical representation of the power intensity is shown.

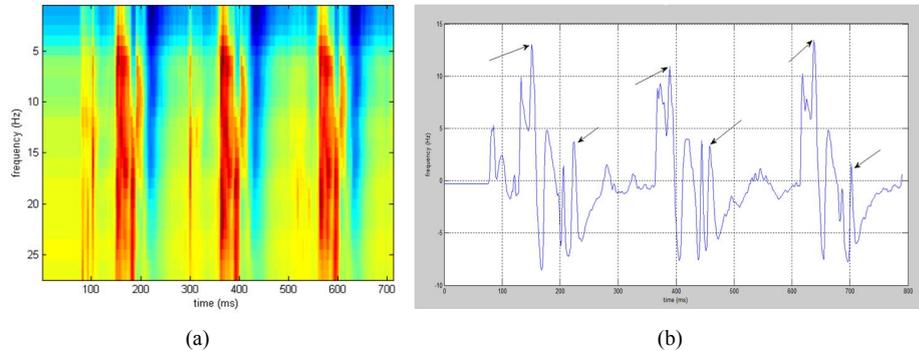

(a)  (b)

Figure 1. (a) The Power spectrum of a speech signal. (b) MFCC of a speech signal of a subject when one word was pronounced for three times. The distance between the indicated peaks are used to form a feature vector.

The Mel-frequency cepstral coefficients (MFCC) are widely used in audio classification experiments due to its good performance. It extracts and represents features of speech signal. The Mel-cepstra takes short-time spectral shape with important data about the quality of voice and production effects [12]. To calculate these coefficients the cosine transform of real logarithm of the short-term spectrum of energy must be done. Then it is performed in mel-frequency scale. Further, after pre-emphasizing the speech segments are windowed (Ajgou et al, 2014). The Hamming window used for this process is a simple window based on reduction of leakage effect. It smears energy from true signal frequency into neighboring ones thus negatively affecting the performance. It also contributes to avoiding the discontinuity of the speech signal in time domain that might occur during Fast Fourier Transform. The concept of windowing is based on multiplying the signal frames by window function w[n] [13].

For emotion classification, we have used a bigger dataset formed using 30 subjects in the age group of 20-45. The subjects consist of an equal proportion of males and females. We recorded the voice of each subject for 30 times. All the recorded data was labelled into three categories/classes of emotions; neutral, anger and joy. Instead of including all of the different human emotions, we have used only a limited number of emotions as it can clearly reveal the fact that there are certain distinguishing emotion related elements in human speech. The prepared dataset had three attributes including feature distance, heart rate and class. The class attribute indicates the emotional state of the person and is labelled as A, B and C for neutral, anger and joy respectively. The data was classifier using weka software. The data was run nine times with the training data percentage varying from 10% to 90% with 10% step. In order to check the effectiveness of the model, the accuracy should not decrease with changing the training set [14]. For a particular training data percentage, the experiment was repeated for 30 times and we chose the average value and estimated the standard deviation. The test for each run performed twice for Accuracy or Percentage Correct and AUC. So, the output of the data mining is an accuracy percentage or AUC value with standard deviations from thirty repetitions for each classifier.

4. Results and Discussion

The effect of emotional state on uttering simple words was studied first. The peak distances were taken for studying the effect. The results plotted for a single subject is shown in Figure.2. The y axes in the graphs indicate the peak to peak distance measure. The graphs shown in Figure.2 are based on the study of sound signals which represent letters only.

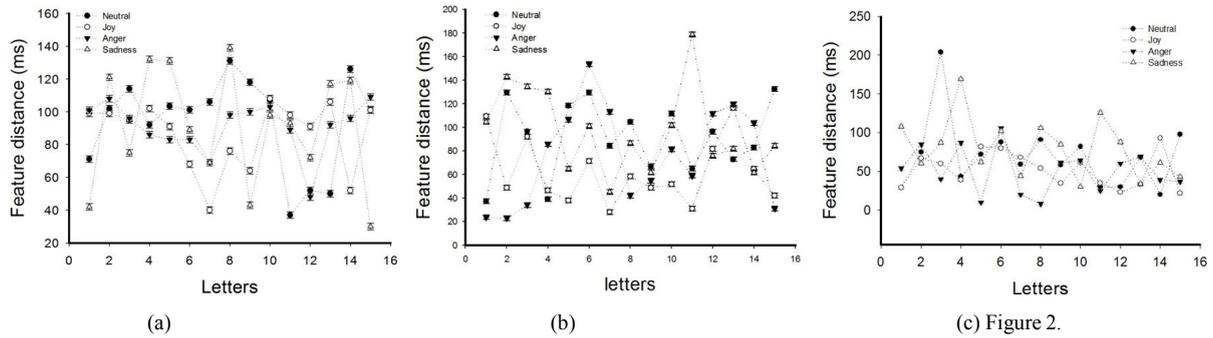

(a)  (b)  (c) Figure 2.

The distances between the peaks for single letters for (a) subject 1, (b) subject 2 and (c) subject 3.

A common behavior can be observed when looking at the graphical results shown in Figure.2(a), 2(b) and 2(c). The features for the emotion sadness have a greater feature distance compared with others. The next higher distance belongs to the neutral emotion. The angry emotional state gives a relatively smaller feature distance and the emotion of joy has the smallest feature distance.

The study of feature distances was done for words also. The graphs obtained were plotted in Figure.3. A similar behaviour was visible here also.

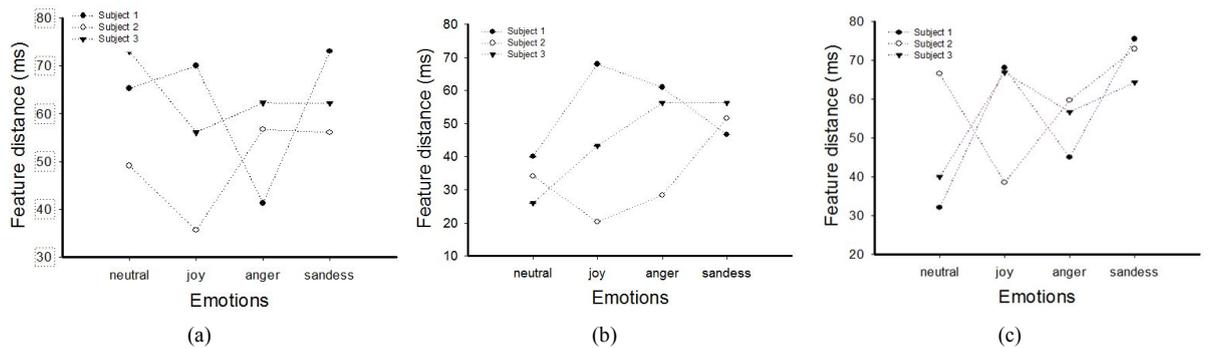

(a)  (b)  (c)

Figure 3. The distances between the peaks of the speech signal of (a) word 1, (b) word 2, and (c) word 3 for all subjects under four emotions: neutral, joy, anger, sadness.

After comparing the pronunciation of words and separate letters contained in those words in Figure.2 and 3, it can be inferred that the emotional state of a person has a direct influence over the vocal signals he/she produces regardless of what is being said.

Now we discuss the results obtained after analysing a group of people rather than few individuals. The main objective is to identify the parameters that influence on the accuracy of the emotion classification. The analysis is based on recognition accuracy, AUC and standard deviation parameters. The accuracy describes how many instances were correctly classified. As it is not sufficient for data analysis due to the fact that it is not sensitive to class distribution and there no chance to make corrections, another parameter AUC, i.e. Area under the ROC curve is also considered. The AUC is a probability of correctly identified classes and it determines quality of a classification model [9]. For more detailed analysis and getting precise results, the standard deviation of these two parameters, which shows variation of the accuracies of thirty repeated runs of each instances from the group mean, were calculated and examined as well.

### 4.1. Recognition Accuracy Analysis

Figure.4 displays the recognition accuracy obtained for seven different classifiers using different training data percentages. The vertical lines in the graph represent the standard deviation of the classifiers accuracy for each train size. Figure.4(a) is plotted for the data collected from all the 30 subjects and Figure.4(b) shows classification accuracy for the data collected from one subject. Both graphs show that the highest accuracy is observed for a training size of 10%, where all classifiers have the smallest range of percentage between approximately 36.5% and 39.5%. While the train size increases this range grows until it reaches the range from about 37% to 45%.

The same result is observed for the accuracy of emotion classification of one subject. However, despite the fact that classifiers are split into two groups: Naive Bayes, Lazy IB1 and RBF Network are in one group, and Logistic, Ada Boost M1, Bagging and Random tree in another one. The Figure 2 presents the same graph but without standard deviation in order to see how the classifiers' accuracies are distributed compared to the Figure.4(a).

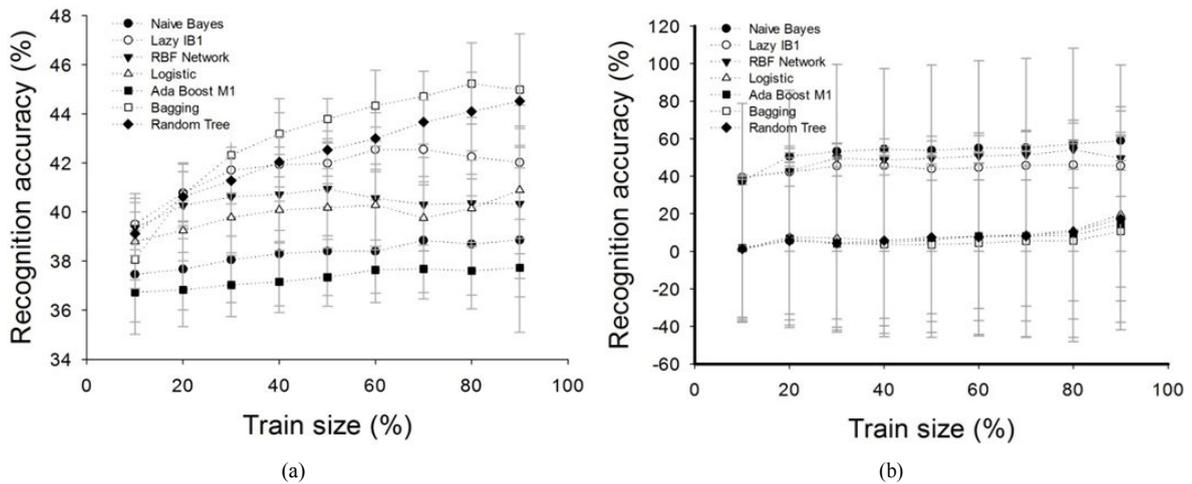

Figure 4. Recognition accuracy of all seven classifiers for different train size, i.e. model, for (a) a group of subjects and (b) an individual subject

### 4.2. AUC Analysis

The behavior of the AUC is similar to the recognition accuracy. The AUC analysis of the individual subject shows better performance, as all classifiers are located closer to each other. Figure.5(a) and 5(b) represents the AUCs for data collected from one subject.

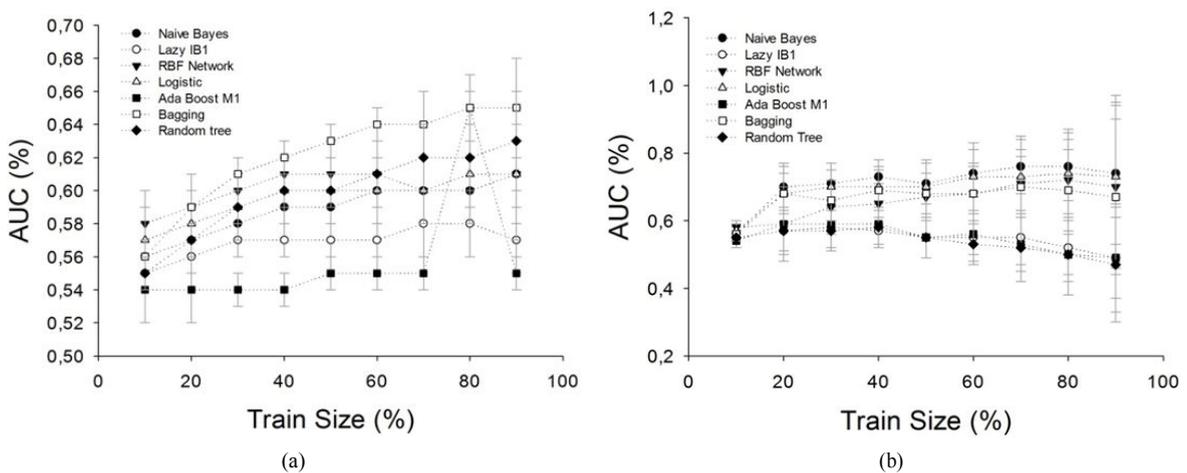

Figure 5. Area under the ROC curve of all seven classifiers for different training size, i.e. model, for (a) a group of subjects and for (b) an individual subject.

From the results obtained, it can be observed that it is more reliable to rely on data from a single subject rather than the data obtained from a group.

## 5. Conclusion

The paper explored the idea of detecting the emotional state of a person by speech processing techniques. The study on words and letters under different emotional situations proved that the emotional state can alter the speech signal. It was observed that there are distinguishable features in a speech segment that characterizes each emotion state. After performing the classification tests on a dataset prepared from 30 subjects, it was observed that it is better considering data from an individual subject rather than a group of people. The development of a software based agent for emotion detection and heart rate analysis can greatly improve telemedicine based systems.